\documentclass[aip,jcp,reprint,numerical,amsmath,amssymb,amsfonts]{revtex4-1}
\usepackage{bm,graphicx,xcolor,hyperref,rotating,lineno,mathtools,multirow,makecell,placeins,physics}
\usepackage[version=4]{mhchem}
\usepackage{hyperref,siunitx}

\hypersetup{
    colorlinks=true,
    linkcolor=blue,
    filecolor=blue,
    urlcolor=blue,
    citecolor=blue
}

\usepackage[normalem]{ulem}

\newcommand{\br}{\mathbf{r}}

\newcommand{\bA}{\mathbf{A}}

\newcommand{{\dno}}{\Delta\text{NO}}
\newcommand{{\dnoof}}{\Delta\text{NO-OF}}
\newcommand{{\dnocs}}{\Delta\text{NO-CS}}

\newcommand{\Mem}{Department of Chemistry, Memorial University, St. John's, Newfoundland and Labrador, A1B 3X7, Canada}
\newcommand{\UMan}{Department of Chemistry, University of Manitoba, Winnipeg, Manitoba, R3T 2N2, Canada}
\newcommand{\UWin}{Department of Chemistry, University of Winnipeg, Winnipeg, Manitoba, R3B 2E9, Canada}

\begin{document}
\preprint{atomicEc}

\title{Atomic Radial Correlation Energy Density Components}
\author{Ibrahim E. Awad}
\affiliation{\Mem}
\author{Abd Al-Aziz A. Abu-Saleh}
\affiliation{\Mem}
\author{Gurleen Cheema}
\affiliation{\UMan}
\author{Joshua W. Hollett}
\email[Corresponding author: ]{j.hollett@uwinnipeg.ca}
\affiliation{\UWin}
\affiliation{\UMan}
\author{Raymond A. Poirier}
\affiliation{\Mem}
\date{\today}

\begin{abstract}
	The components of the radial correlation energy density are calculated and analyzed for the atoms from \ce{He} to \ce{Ar}.  The components include the nucleus-electron potential correlation energy density, the kinetic correlation energy density and the electron-electron potential correlation energy density.  The necessary correlated one and two-electron density matrices are obtained from the Extrapolated-Full-Configuration-Interaction (exFCI) wave function where the reference wave function is restricted Hartree-Fock (RHF) or restricted open-shell Hartree-Fock (ROHF) depending on whether the atom is closed or open-shell. The accuracy associated with integrating the HF and exFCI energy density components, and the correlation energy density components, is evaluated on the SG-1 and SG-2 atomic grids. The SG-1 grid provides atomic energies that are accurate to about 1 kJ mol$^{-1}$, with the exception of the kinetic energy.  The SG-2 grid is required for the analysis of atomic kinetic energies and more subtle energetic effects.  There is also a significant amount of integration error cancellation in the correlation energy densities.  The radial correlation energy densities display notable shell structure, and there is a substantial difference between the $\alpha$ and $\beta$-electron correlation energy densities for the open-shell atoms.
\end{abstract}

\maketitle

\section{Introduction}
\label{sec:intro}

The partitioning of molecules into their constituent atoms, along with the associated energies, is a topic of continual debate and discussion, mainly due to its conceptual ambiguity.\cite{andres2019nine, andrada2020EDApath, cassam2001some, racioppi2020generalized, Pendas2023atoms}  However, the value of such an approach to the understanding of the physical and chemical properties of atoms and molecules has been demonstrated over and over again.\cite{bader1984bonded, bader1984characterization, hopffgarten2012EDA, zhao2018EDANOCV, glendening2005NEDADFT, mei2015num}  The information extracted from such approaches depends on how the electron density is partitioned in real-space, or the wave function is paritioned in Fock-space, in conjunction with the assignment of energy contributions to individual atoms and their interactions.  Therefore, in an effort to gain further understanding of molecular interactions of different types and scales, various approaches to decomposing the electronic energy have been devised in addition to, and in cooperation with, methods for partitioning these energy components into atomic contributions.\cite{morokuma1971EDA, hirshfeld1977bonded, bader1990atoms, angyan1994covalent, glendening1994NEDA, rafat2005atom, blanco2005QTAM, mitoraj2009ETSNOCV, popelier2001Coulomb, vyboishchikov2019iterative, fias2018reference, Heidar2017info, Heidar2024var}  Despite the apparent seamless nature of the molecular wave function, useful data for understanding chemistry can be extracted from the arbitrarily defined atoms within. 

In general, a scheme that partitions a molecule into atoms may do so with the one-electron density according to
\begin{equation}
	\rho^A(\br) = w_A(\br) \rho(\br),
\end{equation}
where $w_A(\br)$ are the atomic weights that assign a contribution of the total one-electron density, $\rho(\br)$, to a particular atom, $A$. This partitioning is a common feature of Energy Decomposition Analysis (EDA). There are many schemes that achieve such partitioning, some of which are quite distinct\cite{hirshfeld1977bonded, bader1990atoms, glendening1994NEDA, Heidar2017info} and offer their own advantages and disadvantages.\cite{Pendas2023atoms}  In most cases the partitioning is fuzzy, meaning the atomic densities overlap within the molecule,\cite{hirshfeld1977bonded, becke1988multicenter} however in the case of Bader's atoms-in-molecules\cite{bader1990atoms} the molecule is cleanly partitioned along zero-flux density surfaces.  

With regards to analyzing atomic densities, it was recently shown that the introduction of the radial Jacobian factor, $4\pi r_A^2$, to the atomic density
\begin{equation} \label{eq:radrho}
	\rho^A_\text{rad}(\br) = 4 \pi r_A^2 \rho^A(\br),
\end{equation}
where $r_A = | \br - \bA |$ is the distance from atom $A$, is particularly useful for the analysis of the shell structure of the atomic densities, in isolation or within a molecule.\cite{Warburton2011,Besaw2015}  The radial atomic density, $\rho^A_\text{rad}(\br)$, can clearly illustrate the deformation of the core and valence density of atoms in molecules compared to isolated atoms.  Also, the $r_A^2$ weighting amplifies the valence electron contribution which, in standard approaches, is often difficult to discern compared to the core contribution.

This radial atomic density approach has also been extended to energy densities\cite{awad2019} in a similar fashion as the grid-EDA approach.\cite{imamura2007grid} Analogous to the one-electron density, radial atomic energy densities are defined for each component of the electronic energy,
\begin{equation} \label{eq:radx}
	\xi^A_\text{rad}(\br) = 4 \pi r_A^2 \xi^A(\br),
\end{equation}
where $\xi^A(\br)$ is the atomic contribution of the particular energy density component, which can be assigned using atomic weights.  Radial atomic energy densities allow for a simplified, one-dimensional, analysis of the influence of interatomic interactions within a molecule on the different components of the electronic energy.  The approach has previously been applied to the Hartree-Fock energy of atoms and molecules\cite{awad2019}.  This study focuses on the post-Hartree-Fock correlation energy density components within individual atoms, by employing the one- and two-electron reduced density matrices of the Hartree-Fock and extrapolated Full Configuration Interaction wave functions.  This allows for the analysis of the effects of correlation on all the components of the electronic energy; kinetic, nucleus-electron potential and electron-electron potential.  Understanding these effects in individual atoms is an essential  step before analyzing the effects of electron correlation on bonding and other molecular interactions.

\section{Theory}
\label{sec:theory}
The electronic energy, $E$, of an atom or molecule is given by the following expression
\begin{equation}
	E = \sum_{ab} \gamma_{ab} \left( t_{ab} + v^\text{ne}_{ab} \right) + \sum_{abcd} \Gamma_{abcd} \langle ab | cd \rangle \;,
\end{equation}
where the one-electron density matrix, $\bm{\gamma}$, and two-electron density matrix, $\mathbf{\Gamma}$, corresponding to the particular orbital basis, $\{ \psi_a(\br) \}$, are employed. The density matrices multiply the appropriate one- and two-electron energy integrals.  These include the kinetic energy integrals,
\begin{equation}
t_{ab} = -\tfrac{1}{2} \int \psi_a^*(\br) \nabla^2 \psi_b(\br) d\br = \tfrac{1}{2}\int \nabla \psi^*_a(\br) \nabla \psi_b(\br) d\br \;,
\end{equation}
the nucleus-electron potential energy integrals for a system of $M$ nuclei,
\begin{equation}
v^\text{ne}_{ab} = \sum_{A=1}^M \int \psi_a^*(\br) \frac{-Z_A}{r_A} \psi_b(\br) d\br \;,
\end{equation}
where $Z_A$ is the atomic number, and the electron-electron repulsion integrals,
\begin{equation}
\langle ab | cd \rangle = \int \int \psi_a^*(\br_1)\psi_b^*(\br_2) r_{12}^{-1} \psi_c(\br_1)\psi_d(\br_2) d\br_1 d\br_2 \;,
\end{equation}
where $r_{12}  = |\br_1 - \br_2|$.

In the case of restricted open-shell Hartree-Fock (ROHF), the 1-RDM is given by,
\begin{equation}
	\gamma_{ab}^\text{ROHF} = f_a \delta_{ab} \;,
\end{equation}
where $f_a$ is the electron occupancy factor, which may have a rational value between 0 and 2.  The ROHF 2-RDM may also be expressed in terms of the occupancy factor,
\begin{equation}
	\Gamma_{abcd}^\text{ROHF} = \frac{f_a f_b}{2} \left( j_{ab} \delta_{ac}\delta_{bd} - k_{ab} \delta_{ad}\delta_{bc} \right)
\end{equation}
where $j_{ab}$ and $k_{ab}$ are the Coulomb and Exchange coupling terms, respectively.\cite{roothaan1960rohf, plakhutin1997rohf, krebs1999rohf, tsuchimochi2010rohf}

Expressions for the local energy density, $\xi(\br)$, components easily follow from the integrals above.  In the case of kinetic energy, two expressions are considered.  The first involves the Laplacian with respect to the electron coordinates,
\begin{equation}
\label{eq:ts}
T^s(\br) = -\tfrac{1}{2} \sum_{ab} \gamma_{ab} \psi_a^*(\br) \nabla^2 \psi_b(\br).
\end{equation}
This definition of local kinetic energy density is referred to as the Schr{\"o}dinger kinetic energy density.  Through use of integration by parts, an alternative expression for the kinetic energy can be realized that is positive definite,
\begin{equation}
\label{eq:tp}
T^+(\br) = \tfrac{1}{2} \sum_{ab} \gamma_{ab} \nabla\psi^*_a(\br) \nabla\psi_b(\br).
\end{equation}
While these are the most obvious expressions for the local kinetic energy density, and have been examined in the past,\cite{imamura2007grid} the quantity does not have a unique definition.\cite{ayers2002, anderson2010}

The potential energy densities have unique expressions, and are given as follows,
\begin{equation}
V^\text{ne}(\br) = -\sum_{A=1}^M \frac{Z_A}{r_A} \rho(\br),
\end{equation}
where
\begin{equation}
\rho(\br) = \sum_{ab} \gamma_{ab} \psi_a^*(\br)\psi_b(\br),
\end{equation}
and
\begin{equation}
\label{eq:vee}
V^\text{ee}(\br) = \sum_{abcd} \Gamma_{abcd} \psi_a^*(\br)\psi_c(\br) \int \frac{\psi^*_b(\br_2) \psi_d(\br_2)}{|\br - \br_2|} d\br_2.
\end{equation}
The intermediate integration employed for the electron-electron potential energy, $V^\text{ee}(\br)$, is often referred to as the pseudospectral method and appears in other contexts.\cite{friesner1985PS}

In the case of an RHF or ROHF wave function, the electron-electron potential energy density may be decomposed into the Coulomb energy density,
\begin{equation}
J(\br) = \sum_{ab} j_{ab} \frac{f_a f_b}{2} \psi_a^*(\br)\psi_a(\br) \int \frac{\psi^*_b(\br_2) \psi_b(\br_2)}{|\br - \br_2|} d\br_2,
\end{equation}
and the exchange energy density, 
\begin{equation}
K(\br) = -\sum_{ab} k_{ab} \frac{f_a f_b}{2} \psi_a^*(\br)\psi_b(\br) \int \frac{\psi^*_b(\br_2) \psi_a(\br_2)}{|\br - \br_2|} d\br_2.
\label{eq:K}
\end{equation}
For any of the energy densities the local effect of correlation can be observed through the correlation energy density,
\begin{equation}
	\xi_c(\br) = \xi(\br) - \xi_\text{HF}(\br),
\end{equation}
where the HF (ROHF or RHF) energy density is subtracted from the exact (or near exact, full configuration interaction) energy density.

Finally, any of the radial energy densities may be decomposed into their spin-resolved components by using the particular spin-resolved component of the 1-RDM,
\begin{equation}
\label{eq:1rdmspin}
	\bm{\gamma} = \bm{\gamma}^\alpha + \bm{\gamma}^\beta,
\end{equation}
or 2-RDM,
\begin{equation}
\label{eq:2rdmspin}
	\mathbf{\Gamma} = \mathbf{\Gamma}^{\alpha\alpha} + \mathbf{\Gamma}^{\beta\beta} + \mathbf{\Gamma}^{\alpha\beta} + \mathbf{\Gamma}^{\beta\alpha}
\end{equation}
where $\alpha$ refers to spin-up electrons and $\beta$ refers to spin-down electrons.  The double-superscript, $\sigma\sigma'$, correspond to the spin of electron 1 and 2, respectively.

\section{Method}
\label{sec:method}
Calculations were performed using the aug-cc-pCVTZ/f basis set\cite{dunning1989a, kendall1992a, peterson2002a, prascher2011a, woon1993a, woon1995a, pritchard2019a, feller1996a, schuchardt2007a} for all atoms except \ce{He}.  The \ce{He} calculations were performed using the aug-cc-pVTZ basis set.\cite{woon1994a} The Restricted Hartree-Fock (RHF) and Restricted Open-shell Hartree-Fock (ROHF) wave functions were determined for each closed-shell and open shell atom, respectively.  The HF wave functions served as the reference wave function for the extrapolated-Full Configuration Interaction (exFCI) energies and wave functions, which are calculated using Quantum Package 2.\cite{qp2}  The exFCI energies, and wave function, are determined using a determinant-driven selected configuration interaction algorithm known as CIPSI (Configuration Interaction using a Perturbative Selection made Iteratively)\cite{Huron1973,Giner2013,Giner2015} and extrapolating to the full configuration interaction (FCI) limit using multireference perturbation theory.\cite{Garniron2017,Loos2018}  The exFCI calculations were performed with a maximum number of determinants set to $10^6$. Note that the multireference perturbation theory correction of exFCI is not included in the wave function and subsequent RDMs that are used to calculate the energy densities analyzed here. However, the perturbative correction is a relatively small percentage of the total correlation energy (see Table S1 of Supplementary Information) and is not expected to have a noticeable qualitative effect on the correlation energy densities.  The decomposed energy densities were calculated from the HF and exFCI RDMs using MUNgauss.\cite{mungauss}

\section{Results}
\label{sec:results}

\subsection{Accuracy}
\label{ssec:acc}
An initial assessment of the accuracy of the SG-1\cite{gill1993} and SG-2\cite{dasgupta2017SG2} integration grids was carried out for the energy components of the atoms \ce{He} to \ce{Ar} (Table \ref{tab:Eerror}) and the correlation energy components (Table \ref{tab:Ecerror}).  The numerically integrated energies are compared to the corresponding analytically computed values with the root-mean-square error and maximum errors reported.  The errors for each component are decomposed into the contributions from 2nd and 3rd-row atoms, where \ce{He} is included with the 2nd-row atoms.

\begin{table*}
	\caption{\label{tab:Eerror} Numerical Integration Error in Radial Energy Density Components}
	\begin{ruledtabular}
	\begin{tabular}{ccS[table-format=.6]S[table-format=.5]cS[table-format=.0]S[table-format=.0]}
			&	& \multicolumn{5}{c}{Grid}	\\
				\cline{3-7}\\
			&	& \multicolumn{2}{c}{SG-1 (m$E_h$)}&	& \multicolumn{2}{c}{SG-2 (n$E_h$)} \\
				\cline{3-4} \cline{6-7}\\
	component	& period$^a$	& RMS	& Max & 	& RMS	& \hspace{1cm} Max \hspace{1cm}\\
	\hline\\
	$T^s$		& 2nd	& 0.15	& 0.33 	\;\;\text{(Li)}&	& 35	& -82 \;\;\text{(F)}	\\
			& 3rd	& 470	& -546 	\;\;\text{(P)}& 	& 4008	& 6516 \;\;\text{(Si)}\\
	$T^+$		& 2nd	& 0.007	& 0.014 	\;\;\text{(Ne)}&	& 5	& 10 \;\;\text{(Ne)}	\\
			& 3rd	& 7	& -13 	\;\;\text{(Mg)}&	& 107	& 258 \;\;\text{(Ar)}	\\
	$V^\text{ne}$	& 2nd	& 0.05	& -0.08 \;\;\text{(Ne)}&	& 5	& 7 \;\;\text{(Be)}	\\
			& 3rd	& 1.0	& -1.4 	\;\;\text{(Mg)}&	& 12	& 17 \;\;\text{(Ar)}	\\
	$V^\text{ee}$	& 2nd & 0.000008\hspace{-2cm} & -0.00002 \;\;\text{(N)}& & 5& -8 \;\;\text{(O)}	\\
			& 3rd	& 0.04	& -0.09 \;\;\text{(Na)}&	& 7	& -13 \;\;\text{(Ar)}	\\
	$J$		& 2nd	& 0.01	& -0.02 \;\;\text{(Ne)}&	& 2	& -4 \;\;\text{(C)}	\\
			& 3rd	& 13	& 23 	\;\;\text{(Na)}&	& 112	& 314 \;\;\text{(Mg)}	\\
	$K$		& 2nd	& 0.01	& 0.02 	\;\;\text{(Ne)}&	& 3	& -5 \;\;\text{(Ne)}	\\
			& 3rd	& 13	& -23 	\;\;\text{(Na)}&	& 114	& -319 \;\;\text{(Mg)}\\
	\end{tabular}
	\end{ruledtabular}
	\flushleft
	\footnotesize
	RMS = root-mean-square error\\
	Max = maximum error with atom specified in ( )\\
	m$E_h$, n$E_h$ = millihartrees, nanohartrees\\
	$^a$ \ce{He} is included with 2nd-period atoms
\end{table*}

The energy components analyzed in Table \ref{tab:Eerror} correspond to the exFCI atomic wave functions, with the exception of the Coulomb, $J$, and exchange, $K$, energies, which are from the RHF and ROHF atomic wave functions.  The error in the numerically integrated energies varies widely depending on the component, with the electron-electron potential energy being the most accurate. It is seen that the Schr{\"o}dinger kinetic energy density, $T^s$, exhibits the most integration error.  This result is not necessarily surprising considering the presence of the second derivative which gives the energy density sharper, and more undulating, features compared to other components. Using the positive definite definition of kinetic energy density, $T^+$, reduces this error by one to two orders of magnitude.  This reduction is presumably due to the presence of just the gradient rather than the Laplacian.  The error in the $J$ and $K$ energy components are relatively large and effectively cancel each other.  This can be explained by the use of a self-interaction-free definition of the Coulomb energy.  That is, there is no double-counting of intraorbital electron-electron repulsion, $J_{aa}$, and hence no self-interaction correction in the exchange energy, $k_{aa} = 0$ [see \eqref{eq:K}].  Such a definition of $J$ and $K$ leads to energy density components that can be non-spherical for atoms, whereas their sum, $J(\br)+K(\br)$, is spherical and is easier to accurately integrate.

For all energy components, the error is drastically reduced upon moving from the SG-1 grid to the SG-2 grid.  This is not terribly surprising considering that the SG-1 grid is based on an unpruned grid with 50 radial shells and 194 angular points per atom, while the SG-2 grid is based on an unpruned grid with 75 radial shells and 302 angular points.\cite{dasgupta2017SG2}  The least improvement is seen for the electron-electron potential energy but this is due to the fact that the error is already quite small for the SG-1 grid and hence little to improve upon.  The only error larger than nanohartrees (n$E_h$) is for the Schr{\"o}dinger kinetic energy density of the 3rd-row atoms (RMS of 4008 n$E_h$, and maximum error of 6516 n$E_h$).  However, compared to the SG-1 grid these errors are almost 5 orders of magnitude smaller. Therefore, depending on the application, energy components calculated numerically ({\it e.g.} atom-in-molecule contributions) using the SG-1 grid may have sufficient accuracy (sub kJ mol$^{-1}$) with the exception of kinetic energy.  In applications where more subtle energy differences are of interest then the SG-2 grid would be required.

\begin{table*}
	\caption{\label{tab:Ecerror} Numerical Integration Error in Radial Correlation Energy Density Components}
	\begin{ruledtabular}
	\begin{tabular}{ccS[table-format=.2]S[table-format=.2]cS[table-format=.0]S[table-format=.0]}
			&	& \multicolumn{5}{c}{Grid}	\\
				\cline{3-7}\\
			&	& \multicolumn{2}{c}{SG-1 ($\mu E_h$)}&	& \multicolumn{2}{c}{SG-2 (n$E_h$)} \\
				\cline{3-4} \cline{6-7}\\
	component	& period$^a$	& RMS	& Max & 	& RMS	& \hspace{1cm} Max \hspace{1cm}\\
	\hline \\
	$T_c^s$		& 2nd	& 0.7	& -2.1 	\;\;\text{(Li)}&	& 8	& -12 \;\;\text{(C)}	\\
			& 3rd	& 147	&  221	\;\;\text{(Al)} & 	& 7	& -11  \;\;\text{(P)}	\\
	$T_c^+$		& 2nd	& 0.1	&  0.3	\;\;\text{(Li)}&	& 7	&  10 \;\;\text{(O)}	\\
			& 3rd	& 5	&  8	\;\;\text{(Si)}&	& 6	&  11 \;\;\text{(Na)}	\\
	$V_c^\text{ne}$	& 2nd	& 0.07	&  0.19	\;\;\text{(Li)}&	& 5	&  8 \;\;\text{(Ne)}	\\
			& 3rd	& 10	&  22	\;\;\text{(Si)}&	& 6	&  12 \;\;\text{(Mg)}	\\
	$V_c^\text{ee}$	& 2nd   & 0.3 	&  0.5	\;\;\text{(Ne)} & 	& 7	&  -14 \;\;\text{(N)}	\\
			& 3rd	& 576	& -1122	\;\;\text{(Na)}&	& 11	&  23 \;\;\text{(Ar)}	\\
	\end{tabular}
	\end{ruledtabular}
	\flushleft
	\footnotesize
	RMS = root-mean-square error\\
	Max = maximum error with atom specified in ( )\\
	m$E_h$, n$E_h$ = millihartrees, nanohartrees\\
	$^a$ \ce{He} is included with 2nd-period atoms
\end{table*}
The errors in the numerically integrated correlation energy components are reported in Table \ref{tab:Ecerror}. Notably, the errors using the SG-1 grid are substantially reduced compared to the total energy components (Table \ref{tab:Eerror}). Hence, there is substantial cancellation of the numerical integration error, particularly for the kinetic energy densities.  The electron-electron potential correlation energy benefits the least from this cancellation of error and as a consequence exhibits the largest error for both the SG-1 and SG-2 grids.

\subsection{Correlation Energy Densities}
\label{ssec:Ecden}
The atomic radial correlation energy densities were analyzed in terms of components, nucleus-electron potential, kinetic, electron-electron potential and the total correlation energy.

\subsubsection{Nucleus-electron potential}
\label{sssec:Vne}
\begin{figure}
	\begin{center}
	\includegraphics[width=0.45\textwidth]{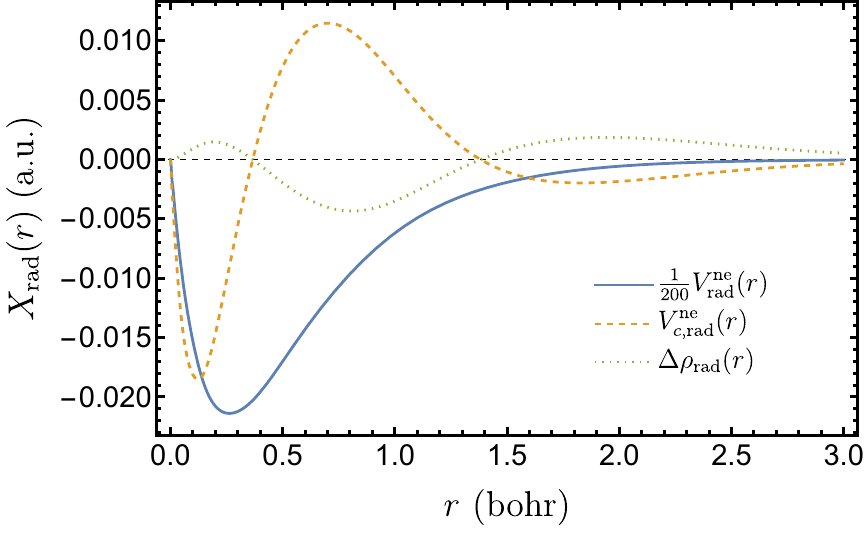}
	\caption{\label{fig:hevne} Radial nucleus-electron potential energy, correlation energy, and change in electron density due to correlation in the \ce{He} atom.}
	\end{center}
\end{figure}

Figure \ref{fig:hevne} compares the radial nucleus-electron energy density (scaled), $V_\text{rad}^\text{ne}(r)$, the radial nucleus-electron correlation energy density, $V_{c,\text{rad}}^\text{ne}(r)$, and the change in the radial one-electron density due to correlation, 
\begin{equation}
	\Delta \rho_\text{rad}(r) = \rho_\text{rad}^\text{exFCI}(r) - \rho_\text{rad}^\text{RHF}(r).
\end{equation}
It is observed that $V_{c,\text{rad}}^\text{ne}(r)$ has a global minimum near the nucleus, $r = 0.12$ bohr, and another minimum far from the nucleus at $r = 1.83$ bohr.  This is explained by the shift in electron density due to the correlated motion of the electrons, specifically radial correlation.\cite{banyard1978mom, koga1996rad, saha2003radang, hollett2011h2}  Radial correlation allows the electrons to alternate positions near and far from the nucleus, which increases their probability of being close or far away from the nucleus at the expense of mid-range distances ({\it i.e.} negative $\Delta \rho_{\text{rad}}(r)$ region).

\begin{figure}
	\begin{center}
	\includegraphics[width=0.45\textwidth]{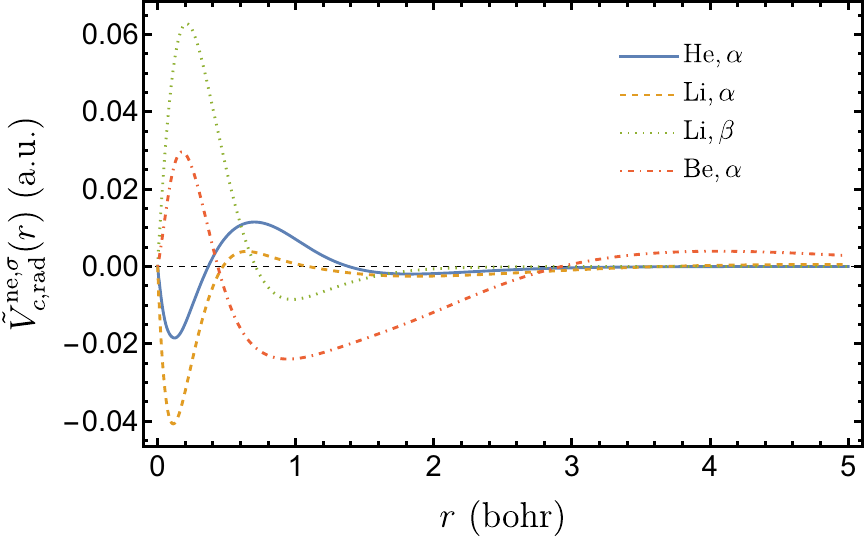}
	\caption{\label{fig:helibevne} Spin-resolved radial nucleus-electron potential correlation energy density, per electron, in \ce{He}, \ce{Li} and \ce{Be} atoms.}
	\end{center}
\end{figure}

In Figure \ref{fig:helibevne}, the spin-resolved components of the nucleus-electron radial correlation energy density of the \ce{He}, \ce{Li} and \ce{Be} atoms are compared.
To allow for a reasonable comparison, the energy densities per electron,
\begin{equation}
	\label{eq:vnepere}
	\tilde{V}^{\text{ne},\sigma}_{c,\text{rad}}(r) = \frac{V^{\text{ne},\sigma}_{c,\text{rad}}(r)}{N^\sigma} \quad \text{(where }\sigma =\alpha\text{ or }\beta\text{)},
\end{equation}
are compared rather than the total.  Note, for closed-shell atoms the $\alpha$ and $\beta$ energy density components are equivalent.  The effect of correlation on the $\alpha$ component of the nucleus-electron potential energy density of \ce{Li} is similar to that of \ce{He}, albeit with a larger decrease close to the nucleus and smaller increase at mid-range.  More interesting, is the near opposite effect of correlation on the $\beta$-energy density, which has a large positive contribution near the nucleus with a small negative contribution at larger distances.  This implies that correlation actually moves $\beta$-electron density away from the nucleus.  The overall effect is almost equal and opposite correlation energy contributions from the $\alpha$ and $\beta$ electrons of \ce{Li}, $V^{\text{ne},\alpha}_c = -0.0222$ $E_h$ and $V^{\text{ne},\beta}_c = 0.0192$ $E_h$. All atomic correlation energy components may be found in the Supplementary Material (Tables S2 and S3).

In the case of \ce{Be}, it is seen that $\tilde{V}^{\text{ne},\alpha}_{c,\text{rad}}(r)$ is positive at both small $r$ and large $r$, indicating a reduction in electron density close-to and far-from the nucleus.  This leads to a subsequent negative region at mid range.

\begin{figure}
	\begin{center}
	\includegraphics[width=0.45\textwidth]{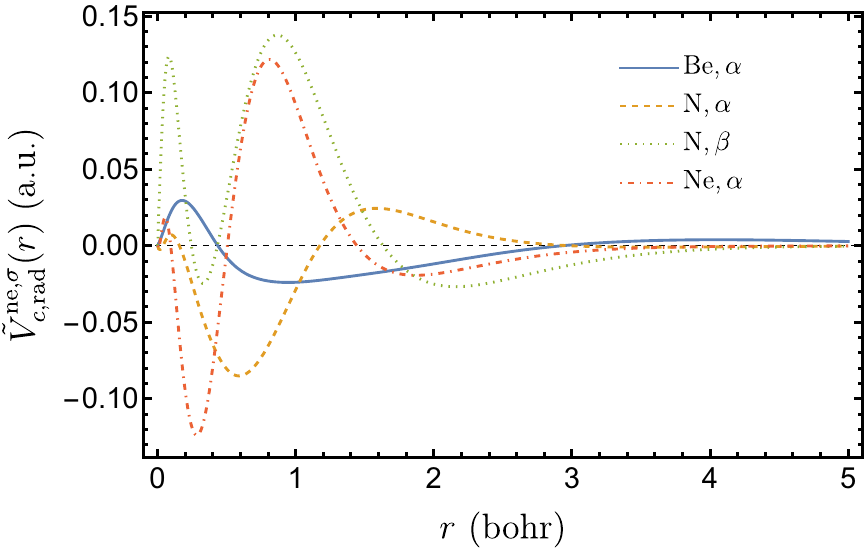}
	\caption{\label{fig:bennevne} Spin-resolved radial nucleus-electron potential correlation energy density, per electron, in \ce{Be}, \ce{N} and \ce{Ne} atoms.}
	\end{center}
\end{figure}

A comparison of the spin-resolved $V^{\text{ne}}_{c,\text{rad}}(r)$ components, per electron, of the \ce{Be}, \ce{N} and \ce{Ne} atoms is shown in Figure \ref{fig:bennevne}.
It is seen that $\tilde{V}^{\text{ne},\alpha}_{c,\text{rad}}(r)$ of \ce{N} shows a reverse effect of correlation compared to \ce{Be}, and is similar to that of \ce{He} with the exception of a small positive region near the nucleus.  The $\beta$-energy density of \ce{N} has a small negative region at small $r$ with large positive contributions on either side, and returns to negative at large $r$.  This depletion of short-range and mid-range $\beta$-electron density actually results in a positive contribution to the correlation energy, $V^{\text{ne},\beta}_c = 0.1460$ $E_h$.

The nucleus-electron radial correlation energy density of the \ce{Ne} atom exhibits features similar to both the $\alpha$ and $\beta$-electron components of the \ce{N} atom.  There is a small positive region and then a negative region at short-range (similar to $V^{\text{ne},\alpha}_{\text{rad}}(r)$ of \ce{N}), followed by a large positive contribution ({\it i.e.} depletion of density) at mid-range and negative contribution at large $r$ (similar to $V^{\text{ne},\beta}_{\text{rad}}(r)$ of \ce{N}).

\begin{figure}
	\begin{center}
	\includegraphics[width=0.45\textwidth]{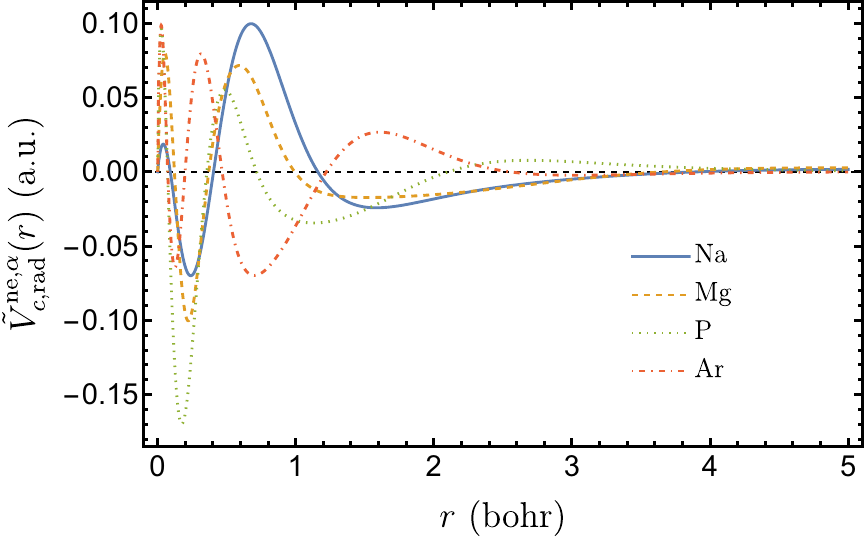}
	\caption{\label{fig:na2aravne} Radial $\alpha$-nucleus-electron potential correlation energy density, per electron, in \ce{Na}, \ce{Mg}, \ce{P} and \ce{Ar} atoms.}
	\end{center}
\end{figure}

The $\alpha$-electron component of $V^{\text{ne}}_{c,\text{rad}}(r)$, per electron, for the \ce{Na}, \ce{Mg}, \ce{P} and \ce{Ar} atoms are compared in Figure \ref{fig:na2aravne}.
It can be seen that the $\alpha$-correlation energy densities of \ce{Na} and \ce{Mg} are similar in shape, however the relative amplitudes of their short and long-range features differ.  The presence of $3p$-electrons, in the \ce{P} and \ce{Ar} atoms, leads to an additional peak ({\it i.e.} positive region) in $V^{\text{ne},\alpha}_{c,\text{rad}}(r)$, located at large $r$.  In the case of the \ce{Ar} atom, the $\alpha$-electron density depletion at large $r$ shifts to smaller $r$ compared to the \ce{P} atom and is accompanied by a density increase, or a region with $V^{\text{ne},\alpha}_{c,\text{rad}}(r) < 0$, at large $r$.

\begin{figure}
	\begin{center}
	\includegraphics[width=0.45\textwidth]{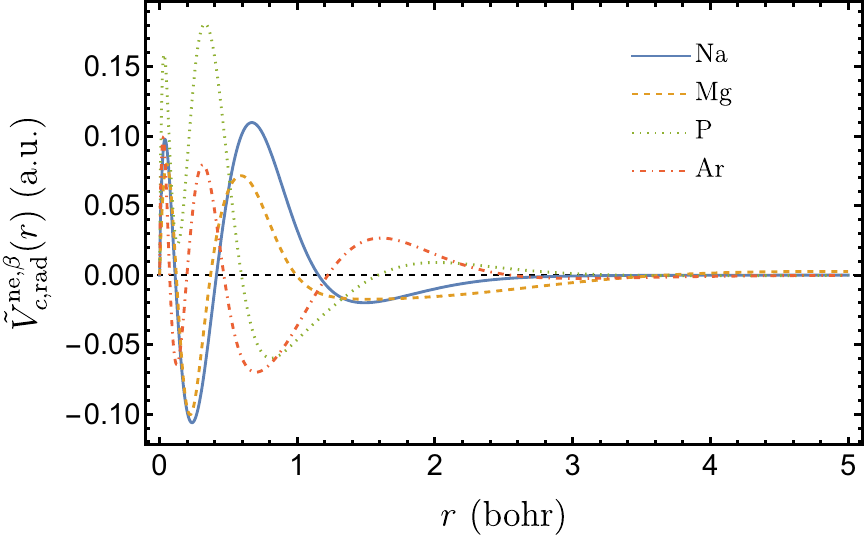}
	\caption{\label{fig:na2arbvne} Radial $\beta$-nucleus-electron potential correlation energy density, per electron, in \ce{Na}, \ce{Mg}, \ce{P} and \ce{Ar} atoms.}
	\end{center}
\end{figure}

In Figure \ref{fig:na2arbvne}, the $V^{\text{ne},\beta}_{c,\text{rad}}(r)$ of the \ce{Na}, \ce{Mg}, \ce{P} and \ce{Ar} atoms are compared.
Again, the radial correlation energy densities of \ce{Na} and \ce{Mg} are quite similar, and actually have closer agreement at small $r$, compared to the $\alpha$-energy densities.  This is not necessarily surprising given that the two atoms have the same number of $\beta$ electrons.  Also, with a shared valence electron configuration, it is perhaps unsurprising that the \ce{P} atom $\beta$-electron energy density closely resembles that of the \ce{N} atom (Figure \ref{fig:bennevne}). And, like the \ce{N} atom, the contribution of this correlation energy component is positive, $V^{\text{ne},\beta}_c = 0.213$ $E_h$.

\subsubsection{Kinetic}
\label{sssec:T}
\begin{figure}
	\begin{center}
	\includegraphics[width=0.45\textwidth]{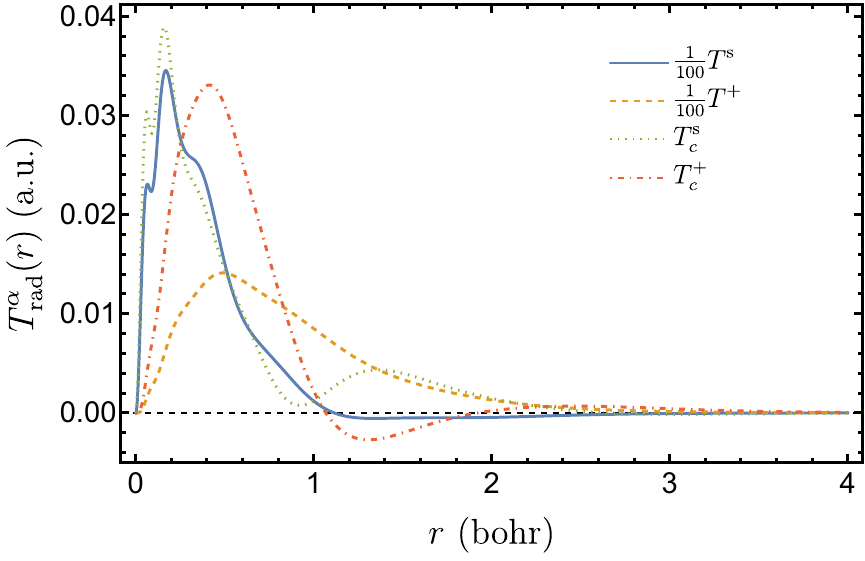}
	\caption{\label{fig:hetstp} Radial kinetic energy density and correlation energy densities in \ce{He}.}
	\end{center}
\end{figure}

Both definitions of the exFCI radial kinetic energy density, scaled by $\tfrac{1}{100}$, and the corresponding kinetic correlation energy densities of the \ce{He} atom are shown in Figure \ref{fig:hetstp}. 
As expected the positive-definite definition of the local kinetic energy, $T^+$ \eqref{eq:tp}, which involves the square-of-the-gradient of the wave function, results in a simpler (less features) radial energy density than the Schr{\"o}dinger definition, $T^\text{s}$ \eqref{eq:ts}, which involves the Laplacian.  These features, or lack thereof, are then echoed in the radial correlation energy densities. It is also seen that, as per the definition, $T^{+,\alpha}_\text{rad}(r) > 0$ for all $r$, whereas $T^{\text{s},\alpha}_\text{rad}(r)$ is negative beyond $r = 1.11$ bohr. Interestingly, the Schr{\"o}dinger definition of the local kinetic correlation energy density is everywhere positive, $T^{s,\alpha}_{c,\text{rad}}(r) > 0$ for all $r$, while $T^{+,\alpha}_{c,\text{rad}}(r)$ is negative for mid-range $r$.

\begin{figure}
	\begin{center}
	\includegraphics[width=0.45\textwidth]{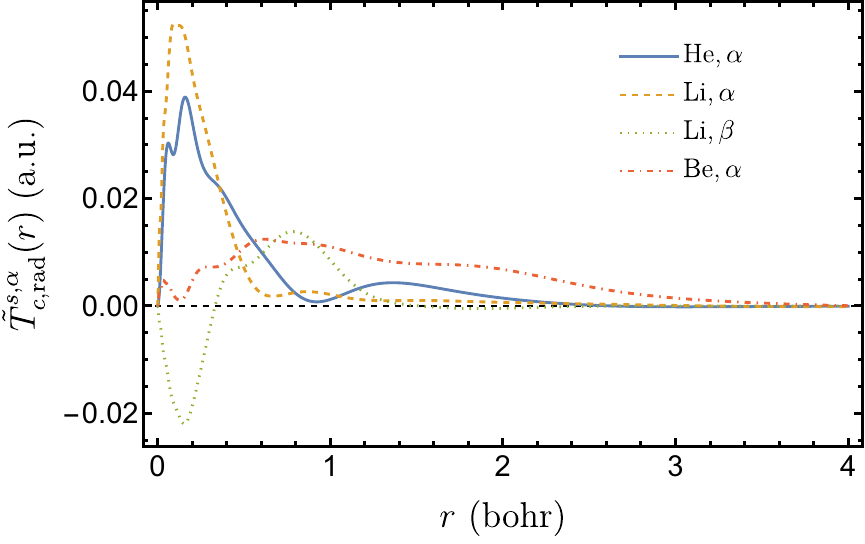}
	\caption{\label{fig:helibets} Radial Schr{\"o}dinger kinetic correlation energy densities, per electron, of \ce{He}, \ce{Li} and \ce{Be}.}
	\end{center}
\end{figure}

The radial Schr{\"o}dinger kinetic correlation energy densities of the \ce{He}, \ce{Li} and \ce{Be} atoms, per electron, are compared in Figure \ref{fig:helibets}.
The $\alpha$-electron kinetic correlation energy densities of \ce{He} and \ce{Li} are both positive everywhere, but there are significant differences in their shape, with the \ce{He} atom density having more pronounced features.  This is likely due to the summation of $1s$-electron and $2s$-electron kinetic energy densities of the \ce{Li} atom smoothing out the features.  The $\beta$-electron kinetic correlation energy density of the \ce{Li} atom differs significantly from the $\alpha$-electron kinetic correlation energy density.  This is due to the depletion of $\beta$-electron density from near the nucleus due to electron correlation, as seen for the nucleus-electron potential correlation energy density.  Like the \ce{He} atom, and the $\alpha$-electrons of the \ce{Li} atom, the Schr{\"o}dinger kinetic correlation energy density of the \ce{Be} atom is positive everywhere, with larger contributions in the mid-range rather than at small $r$.

\begin{figure}
	\begin{center}
	\includegraphics[width=0.45\textwidth]{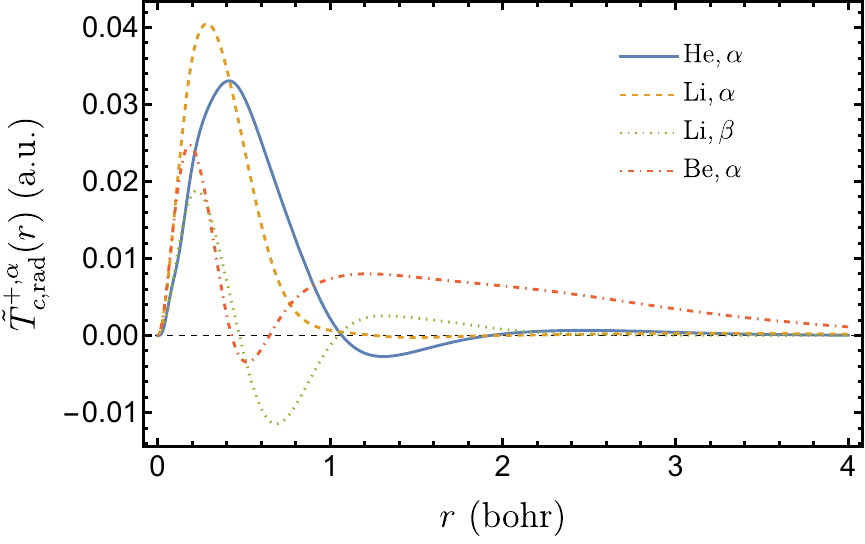}
	\caption{\label{fig:helibetp} Radial positive-definite kinetic correlation energy densities, per electron, of \ce{He}, \ce{Li} and \ce{Be}.}
	\end{center}
\end{figure}

The positive-definite radial kinetic correlation energy densities, per electron, of the \ce{He}, \ce{Li} and \ce{Be} atoms are compared in Figure \ref{fig:helibetp}.
The positive-definite kinetic correlation energy densities are notably simpler than the Schr{\"o}dinger densities, and they all have some region of negative kinetic correlation energy density.  The \ce{He} density and \ce{Li} $\alpha$-density are again quite similar to each other.  In the case of the \ce{Li} $\beta$-density and \ce{Be} density, their shapes differ substantially from the corresponding Schr{\"o}dinger densities, particularly near the nucleus.  The $T^{+,\alpha}_{c,\text{rad}}(r)$ of the \ce{Be} atom has a large peak near the nucleus, unlike $T^{s,\alpha}_{c,\text{rad}}(r)$. The $T^{+,\beta}_{c,\text{rad}}(r)$ of the \ce{Li} atom also has a peak near the nucleus which is opposite to the dip that is seen for the Schr{\"o}dinger density.

\begin{figure}
	\begin{center}
	\includegraphics[width=0.45\textwidth]{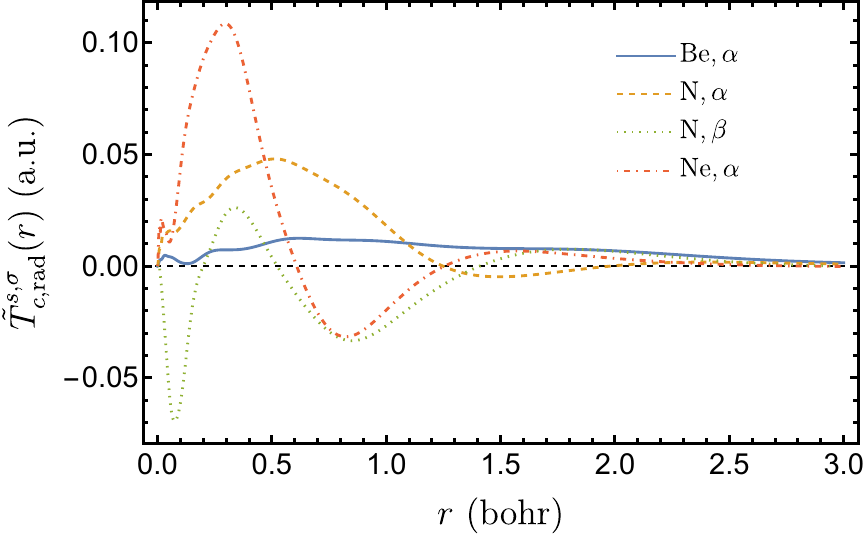}
	\caption{\label{fig:bennets} Radial Schr{\"o}dinger kinetic correlation energy densities, per electron, of \ce{Be}, \ce{N} and \ce{Ne}.}
	\end{center}
\end{figure}

The radial Schr{\"o}dinger kinetic correlation energy densities, per electron, of \ce{Be}, \ce{N} and \ce{Ne} are compared in Figure \ref{fig:bennets}.
The $T^{s,\alpha}_{c,\text{rad}}(r)$ of the \ce{N} atom has a similar shape to that of the \ce{Be} atom, however there is a much larger correlation effect (per-electron), and the correlation energy density has a negative region at larger $r$.  The $\beta$-radial kinetic correlation energy density of the \ce{N} atom has a structure similar to that of the \ce{Li} atom near the nucleus, but also has a negative region at mid-range due to the correlation of the $2p$-electrons.  The additional $\beta$-$2p$-electrons of the \ce{Ne} atom lead to a radial kinetic correlation energy density that is only positive at small $r$, a small peak at very small $r$ and a much larger peak at larger $r$.  Similar to the \ce{N} atom, there is a negative region at mid-range and then a positive contribution at large $r$.

\begin{figure}
	\begin{center}
	\includegraphics[width=0.45\textwidth]{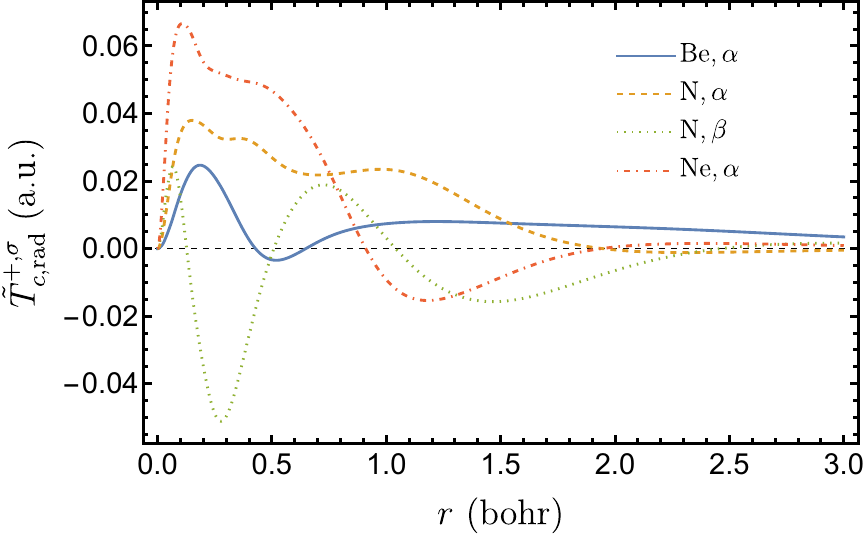}
	\caption{\label{fig:bennetp} Radial positive-definite kinetic correlation energy densities, per electron, of \ce{Be}, \ce{N} and \ce{Ne}.}
	\end{center}
\end{figure}

Figure \ref{fig:bennetp} shows the $\tilde{T}^{+,\alpha}_{c,\text{rad}}(r)$ for the \ce{Be}, \ce{N} and \ce{Ne} atoms.
The radial positive-definite kinetic correlation energy densities of the \ce{N} and \ce{Ne} atoms differ significantly from that of the \ce{Be} atom.  The shape of the \ce{N} atom $\alpha$-energy density is quite similar to that of \ce{Ne} at small $r$, although their amplitudes per-electron differ significantly.  However, at mid-range the $\alpha$-energy density of \ce{N} is positive whereas for \ce{Ne} it is negative.

\begin{figure}
	\begin{center}
	\includegraphics[width=0.45\textwidth]{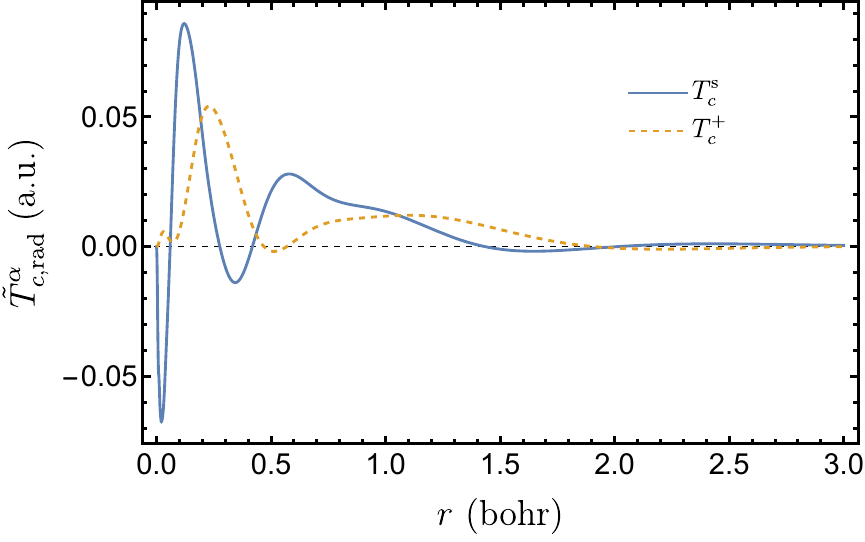}
	\caption{\label{fig:art} Radial kinetic correlation energy densities, per electron, of \ce{Ar}.}
	\end{center}
\end{figure}

The radial Schr{\"o}dinger, and positive-definite, kinetic correlation energy densities of the \ce{Ar} atom are compared in Figure \ref{fig:art}.
The $T^{s,\alpha}_{c,\text{rad}}(r)$ has larger oscillations with $r$, including a sharp dip at small $r$ whereas $T^{+,\alpha}_{c,\text{rad}}(r)$ has a small peak.  The remaining features are similar, however their locations, with respect to $r$, are shifted relative to each other.

\subsubsection{Electron-electron potential}
\label{sssec:Vee}
\begin{figure}
	\begin{center}
	\includegraphics[width=0.45\textwidth]{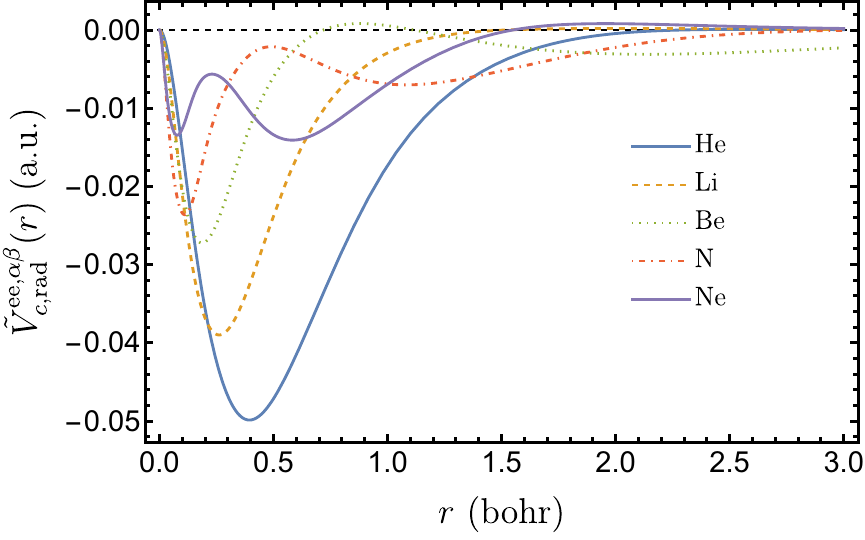}
	\caption{\label{fig:hetoneveeab} Opposite-spin radial electron-electron potential correlation energy densities, per electron pair, of \ce{He}, \ce{Li}, \ce{Be}, \ce{N} and \ce{Ne}}
	\end{center}
\end{figure}

For comparison between atoms, the radial electron-electron potential correlation energy densities are scaled per electron pair; for opposite-spin pairs,
\begin{equation}
	\label{eq:veeopppere}
	\tilde{V}^{\text{ee},\alpha\beta}_{c,\text{rad}}(r) = \frac{V^{\text{ee},\alpha\beta}_{c,\text{rad}}(r)}{N^\alpha N^\beta},
\end{equation}
and for parallel-spin pairs,
\begin{equation}
	\label{eq:veeparpere}
	\tilde{V}^{\text{ee},\sigma\sigma}_{c,\text{rad}}(r) = \frac{V^{\text{ee},\sigma\sigma}_{c,\text{rad}}(r)}{\tfrac{1}{2}N^\sigma \left(N^\sigma - 1\right)}.
\end{equation}
The opposite-spin radial electron-electron potential correlation energy densities, per electron pair, of the \ce{He}, \ce{Li}, \ce{Be}, \ce{N} and \ce{Ne} atoms are compared in Figure \ref{fig:hetoneveeab}. 
The radial electron-electron potential correlation energy density of the \ce{He} atom has the largest magnitude per-electron-pair, with a distribution that roughly follows the shape of the radial electron density, $\rho_\text{rad}(r)$, but shifted to smaller $r$.  The minimum of $V^{\text{ee},\alpha\beta}_{c,\text{rad}}(r)$ for the \ce{He} atom occurs at $r = 0.40$ bohr, whereas the maximum of the exFCI electron density is at $r = 0.57$ bohr.  The correlation energy density of the \ce{Li} atom is similar to that of \ce{He} but with decreased magnitude and shifted to smaller $r$.  The density for the \ce{Be} atom has a positive region, from $r = 0.72$ bohr to $r = 1.12$ bohr, which implies an increase in repulsion between opposite-spin electrons due to correlation. The \ce{Li} and \ce{Ne} atoms also have regions of $V^{\text{ee},\alpha\beta}_{c,\text{rad}}(r) > 0$ at large $r$.  Additionally, a subshell structure appears upon the introduction of $2p$-electrons.  Both the \ce{N} and \ce{Ne} atoms have a second minimum in their correlation energy densities at larger $r$.  The \ce{Be} atom correlation energy density also has a secondary minimum, that occurs at a substantially larger $r$ value and is likely due to its well-known multireference character ({\it i.e.}~contribution from the $2p^2$ valence electron configuration).\cite{hollett2011faces}

\begin{figure}
	\begin{center}
	\includegraphics[width=0.45\textwidth]{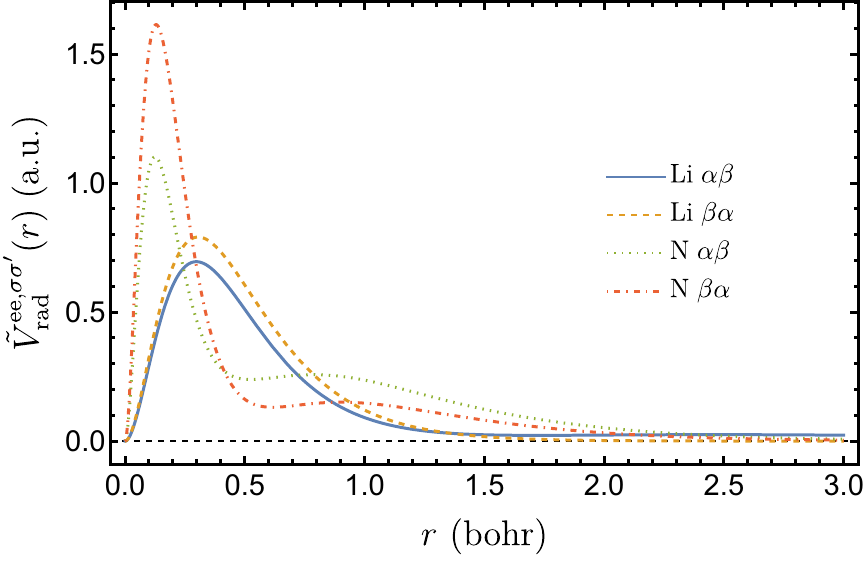}
	\caption{\label{fig:veeabba} Opposite-spin radial electron-electron potential energy densities, per electron pair, of \ce{Li} and \ce{N}.}
	\end{center}
\end{figure}

\begin{figure}
	\begin{center}
	\includegraphics[width=0.45\textwidth]{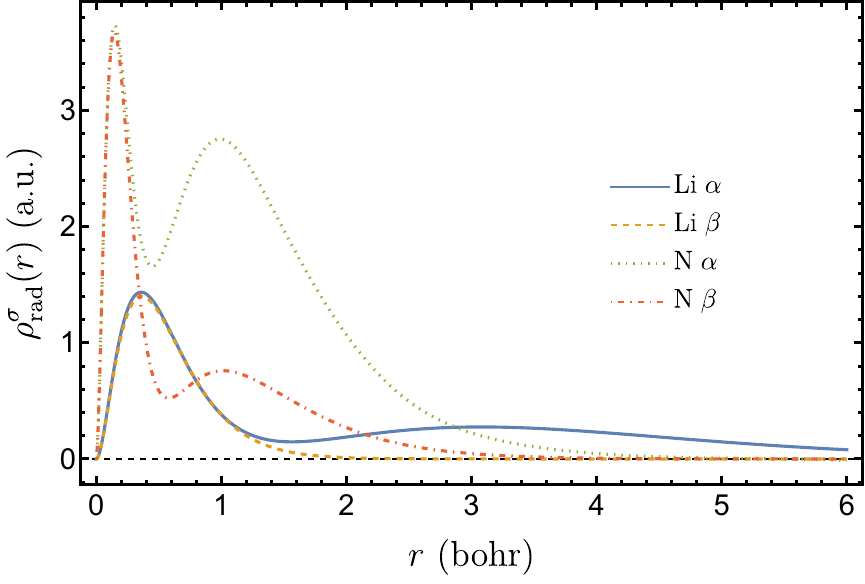}
	\caption{\label{fig:linrhoab} Radial spin-densities of \ce{Li} and \ce{N}.}
	\end{center}
\end{figure}

The opposite-spin radial electron-electron potential energy densities, per electron pair, for the \ce{Li} and \ce{N} atoms are shown in Figure \ref{fig:veeabba}.
For open-shell atoms, $V^{\text{ee},\alpha\beta}_{\text{rad}}(r) \ne V^{\text{ee},\beta\alpha}_{\text{rad}}(r)$.  As per Equations \eqref{eq:vee} and \eqref{eq:2rdmspin}, the coordinate $r$ corresponds to an electron with spin corresponding to the first superscript and the second superscript defines the spin of the electrons that generate the potential.  Therefore, in the case of both atoms,  $V^{\text{ee},\beta\alpha}_{\text{rad}}(r) > V^{\text{ee},\alpha\beta}_{\text{rad}}(r)$ for small $r$ mainly because there are more $\alpha$ electrons for the $\beta$ electron to repel.  At mid-range $r$, $V^{\text{ee},\beta\alpha}_{\text{rad}}(r) \to 0$ because the $\beta$-electron is unlikely to be found that far from the nucleus, whereas $V^{\text{ee},\alpha\beta}_{\text{rad}}(r)$ remains non-zero due to the fact that there are $\alpha$-electrons in the outer valence shell. This is corroborated through analysis of the spin-density, Figure \ref{fig:linrhoab}. 

For the \ce{N} atom, a similar situation to the \ce{Li} atom is observed with respect to the spin-resolved electron-electron potential energy densities.  Near the nucleus, $V^{\text{ee},\beta\alpha}_{\text{rad}}(r) > V^{\text{ee},\beta\alpha}_{\text{rad}}(r)$, because although the $\alpha$ and $\beta$-electron densities are equivalent in this region (see Figure \ref{fig:linrhoab}) the $\beta$ electron also feels the potential from the $2p$ $\alpha$ electrons.  At larger $r$, it is seen that there is a non-zero contribution from $V^{\text{ee},\beta\alpha}_{\text{rad}}(r)$ due to the fact that there is some $\beta$-electron density in the outer valence shell, but it is substantially less than the $\alpha$-electron density. 

\begin{figure}
	\begin{center}
	\includegraphics[width=0.45\textwidth]{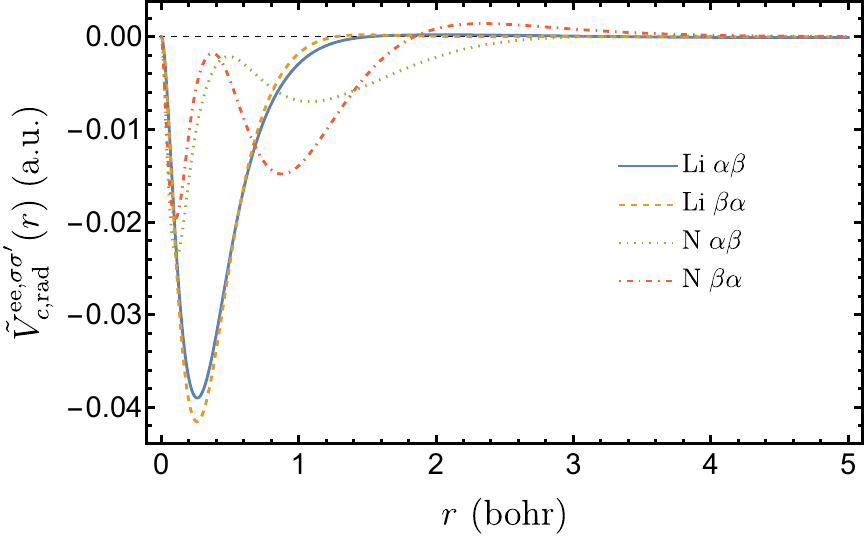}
	\caption{\label{fig:veecabba} Opposite-spin radial electron-electron potential correlation energy densities, per electron pair, of \ce{Li} and \ce{N}.}
	\end{center}
\end{figure}

Given the difference between the $\alpha\beta$ and $\beta\alpha$-components of the radial electron-electron potential energy density, the effects of correlation on both of these components are compared in Figure \ref{fig:veecabba}.
The effects of correlation on $V^{\text{ee},\sigma\sigma'}_{\text{rad}}(r)$ of the \ce{Li} atom are quite similar, with $V^{\text{ee},\beta\alpha}_{c,\text{rad}}(r)$ being more negative near the nucleus and then decaying more quickly to zero.  This closely corresponds to the magnitude of the total electron-electron potential energy density of the \ce{Li} atom, discussed above.  In the case of the \ce{N} atom, the largest difference between the $\alpha\beta$ and $\beta\alpha$-electron-electron potential correlation energy densities is at mid and long-range $r$.  The $\tilde{V}^{\text{ee},\beta\alpha}_{c,\text{rad}}(r)$ has a significantly lower minimum, at $r = 0.88$ bohr, and then becomes positive at large $r$ before decaying.

\begin{figure}
	\begin{center}
	\includegraphics[width=0.45\textwidth]{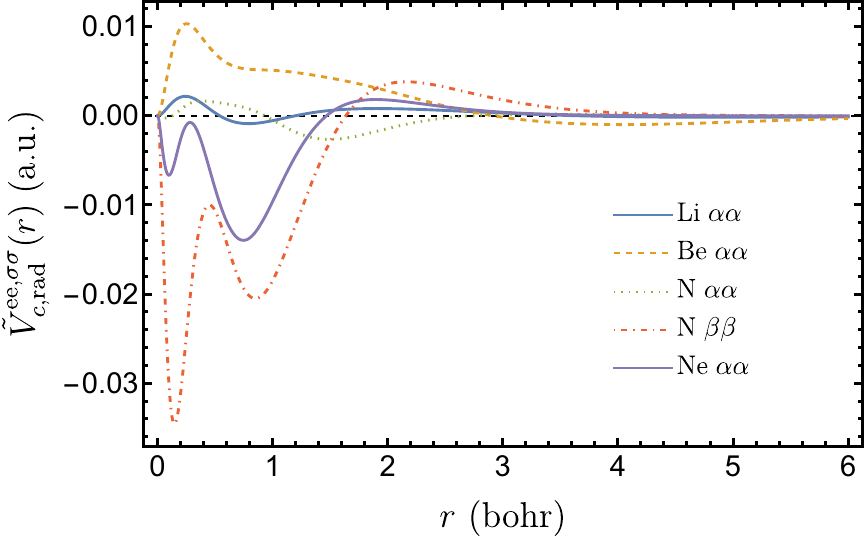}
	\caption{\label{fig:litoneveess} Parallel-spin radial electron-electron potential correlation energy densities, per electron pair, of \ce{Li}, \ce{Be}, \ce{N} and \ce{Ne}.}
	\end{center}
\end{figure}

The parallel-spin radial electron-electron potential correlation energy densities, per electron pair, of the \ce{Li}, \ce{Be}, \ce{N} and \ce{Ne} atoms are compared in Figure \ref{fig:litoneveess}.
The $V^{\text{ee},\alpha\alpha}_{c,\text{rad}}(r)$ of \ce{Li} is positive at short and long-range, with a shallow negative region at mid range.  The result is a slight increase in repulsion and a small positive contribution to the total correlation energy, $V^{\text{ee},\alpha\alpha}_{c}= 0.000977$ $E_h$. In the case of the \ce{Be} atom, the effect of correlation is more significant.  At short and mid-range,  $V^{\text{ee},\alpha\alpha}_{c,\text{rad}}(r)$ is positive and then becomes negative only at long range.  The positive contribution to the correlation energy is an order of magnitude larger, $V^{\text{ee},\alpha\alpha}_{c}= 0.009475$ $E_h$, than that seen in the \ce{Li} atom.  The effect of correlation on $\alpha\alpha$-electron pairs of the \ce{N} atom is qualitatively opposite to that seen in the \ce{Li} atom.  At very small $r$, $V^{\text{ee},\alpha\alpha}_{c,\text{rad}}(r)$ of \ce{N} is negative and then positive at larger $r$ before turning negative at even larger $r$.  However, these effects are small compared to the magnitude of  $V^{\text{ee},\beta\beta}_{c,\text{rad}}(r)$ for the \ce{N} atom.  Two large dips are seen at short to mid-range $r$, followed by a peak at larger $r$.  Electron correlation reduces $\beta\beta$-electron pair interactions near the nucleus (between the $1s$ and $2s$ electrons) and moves them further away.  The effects of correlation on the electron-electron potential energy between parallel-spin electrons in the \ce{Ne} atom closely resembles the effects on $\beta\beta$-electron pairs in the \ce{N} atom, however the relative magnitude of the inner and outer dips is reversed.  The overall magnitude of the effects in the \ce{Ne} atom, per electron-pair, is also reduced compared to the \ce{N} atom.

\begin{figure}
	\begin{center}
	\includegraphics[width=0.45\textwidth]{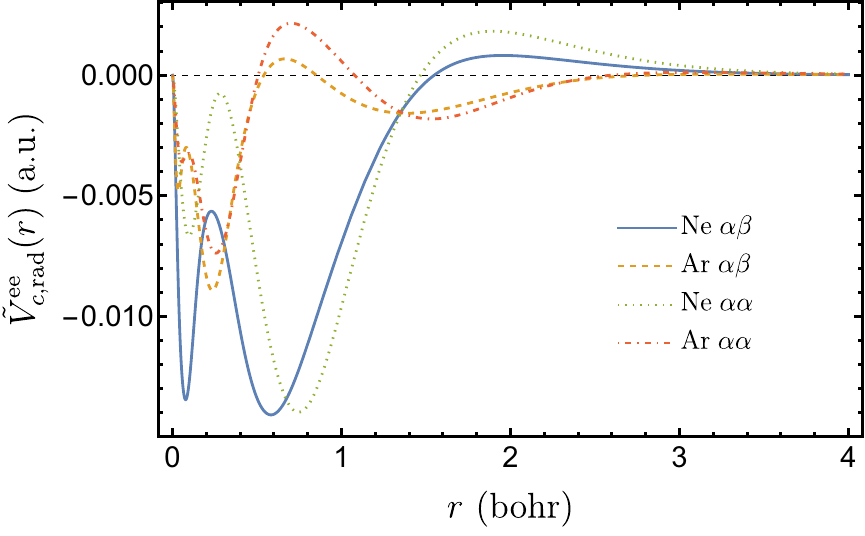}
	\caption{\label{fig:nearvee} Radial electron-electron potential correlation energy densities, per electron pair, of \ce{Ne} and \ce{Ar}.}
	\end{center}
\end{figure}

The opposite and parallel-spin radial electron-electron potential correlation energy densities, per electron pair, for the \ce{Ne} and \ce{Ar} atoms are compared in Figure \ref{fig:nearvee}.
The most striking observation, is the similarity between the opposite-spin and parallel-spin correlation energy densities. The $\alpha\beta$ and $\alpha\alpha$ correlation energy densities all have the same general features, with some variation in the dip and peak magnitude and location.  The magnitude of correlation at short-range is substantially larger for parallel-spin electrons of the \ce{Ne} atom than that of opposite-spin electrons.  In the case of the \ce{Ar} atom, the largest deviation between the opposite-spin and parallel-spin correlation energy densities is in the positive region at mid-range.  As seen with the \ce{Ne} atom, correlation increases the repulsion between parallel-spin electrons at mid-range more than it does for opposite-spin electrons.  

\subsubsection{Total Correlation Energy and the Virial Theorem}
\label{sssec:ec}

\begin{figure}
	\begin{center}
	\includegraphics[width=0.45\textwidth]{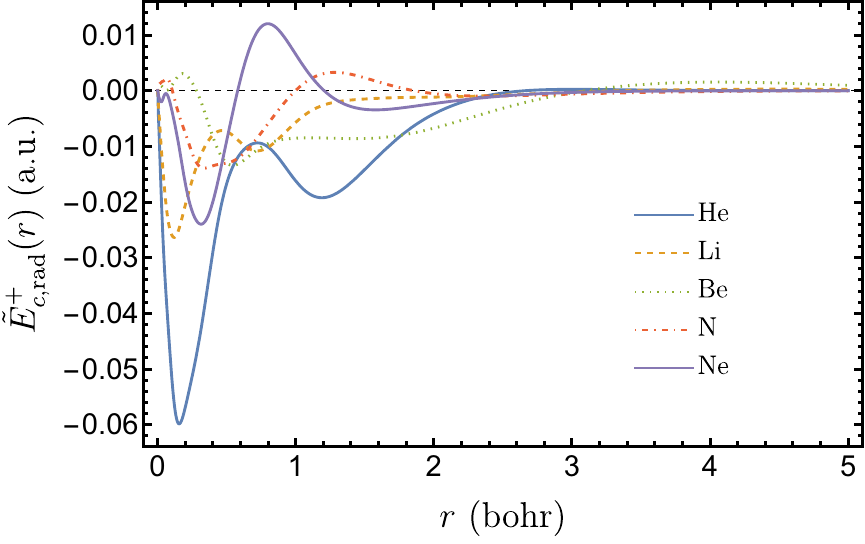}
	\caption{\label{fig:hetoneecp} Radial total correlation energy densities, per electron pair, of \ce{He} and second row atoms [where $T_c(r) = T_c^+(r)$]}
	\end{center}
\end{figure}

The total radial correlation energy densities, per electron pair, of the \ce{He}, \ce{Li}, \ce{Be}, \ce{N} and \ce{Ne} atoms are compared in Figure \ref{fig:hetoneecp}.
The total energy densities in Figure \ref{fig:hetoneecp} are those defined in terms of the positive-definite kinetic energy density,
\begin{equation}
	\label{eq:Ecp}
	E^+_{c,\text{rad}}(r) = T^+_{c,\text{rad}}(r) + V^\text{ne}_{c,\text{rad}}(r) + V^\text{ee}_{c,\text{rad}}(r),
\end{equation}
and
\begin{equation}
	\tilde{E}^+_{c,\text{rad}}(r) = \frac{E^+_{c,\text{rad}}(r)}{\tfrac{1}{2}N\left(N-1\right)},
\end{equation}
where $N$ is the total number of electrons. It is seen that, per-electron pair, the correlation energy density of \ce{He} has the largest magnitude.  There may be slight bias amongst the atoms due to basis set incompleteness, however the primary cause is the abundance of spatial freedom afforded the electrons of the \ce{He} atom compared to the atoms of the second row, particularly those later in the period.\cite{davidson1991, chakravorty1993, chakravorty1996, todd2021mbatoms}  Another notable observation, is the appearance of positive regions of the total correlation energy density for the \ce{Be} atom and beyond.  There is a small positive peak in the \ce{Be} atom correlation energy density at small $r$ and then a wider but shallow positive contribution at large $r$.  In the case of the \ce{Ne} atom, there is one positive region which occurs at mid-range with a peak around $r = 0.81$ bohr.

\begin{figure}
	\begin{center}
	\includegraphics[width=0.45\textwidth]{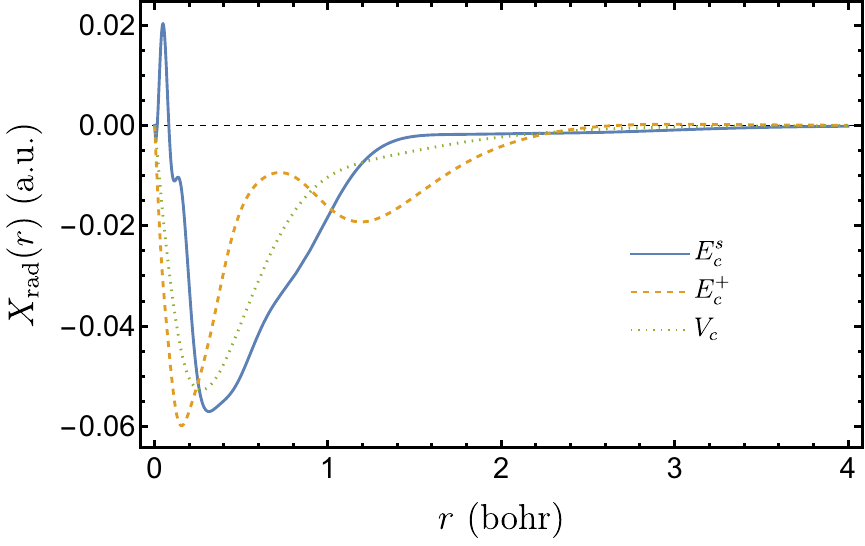}
	\caption{\label{fig:hetv} Radial correlation energy densities and potential correlation energy densities of \ce{He}.}
	\end{center}
\end{figure}

\begin{figure}
	\begin{center}
	\includegraphics[width=0.45\textwidth]{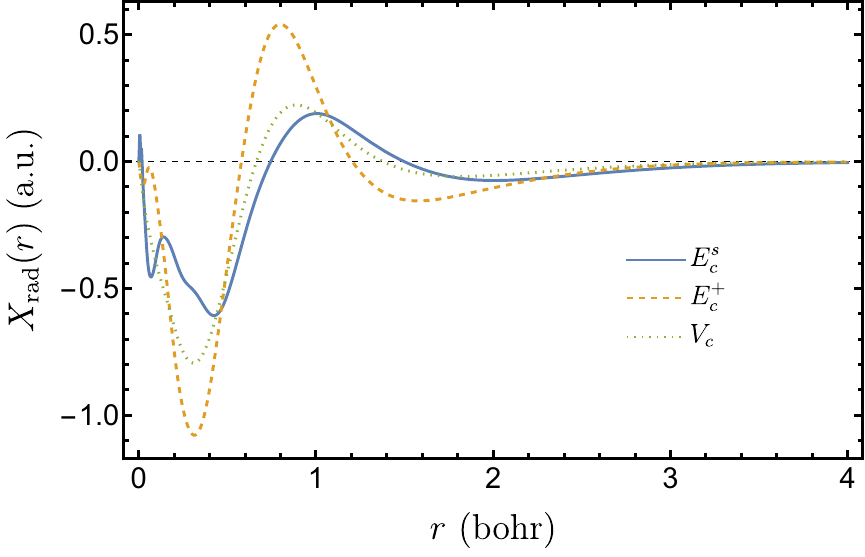}
	\caption{\label{fig:netv} Radial correlation energy densities and potential correlation energy densities of \ce{Ne}.}
	\end{center}
\end{figure}

\begin{figure}
	\begin{center}
	\includegraphics[width=0.45\textwidth]{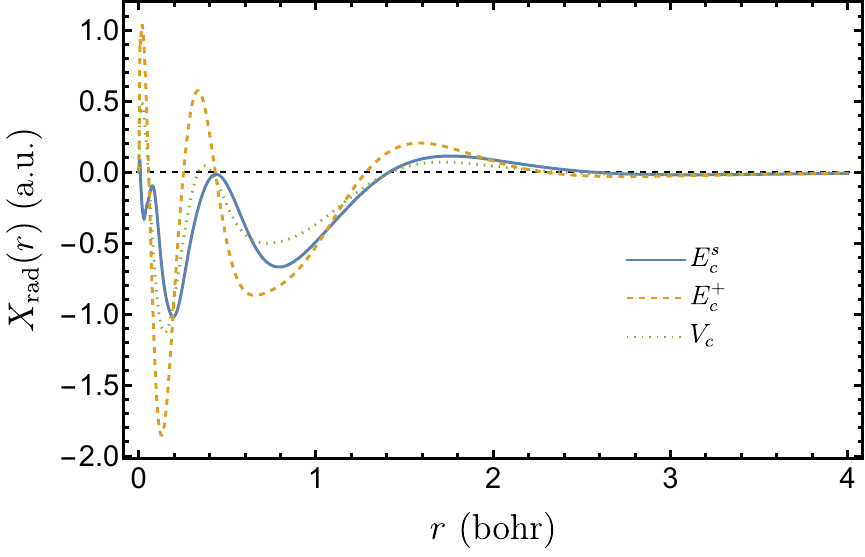}
	\caption{\label{fig:artv} Radial correlation energy densities and potential correlation energy densities of \ce{Ar}.}
	\end{center}
\end{figure}

The remaining three Figures, Figures \ref{fig:hetv}, \ref{fig:netv} and \ref{fig:artv}, provide a comparison of the total correlation energy and the total potential correlation energy to evaluate the locality of the Virial Theorem.
The Virial Theorem, applied to the correlation energy of atoms, gives
\begin{equation}
\label{eq:vt}
	2 T_c = -V_c
\end{equation}
where $V_c$ is the total potential correlation energy, $V_c  = V^\text{ne}_c + V^\text{ee}_c$.  By using the equation, $E_c = T_c + V_c$, the following relation between the potential and total correlation energies is found,
\begin{equation}
\label{eq:ev}
	E_c = \tfrac{1}{2}V_c,
\end{equation}
The Virial Theorem does not imply agreement on a local basis.  That is, $E_c(r) = \tfrac{1}{2}V_c(r)$ is not guaranteed, nor likely, to be satisfied.  This is particularly obvious when considering that there is no unique definition of the local kinetic energy density.\cite{ayers2002, anderson2010}  However, considering the approaches used in density functional approximation development and related energy decomposition analysis, an analysis of the agreement, or lack thereof, with the Virial Theorem on a local basis is enlightening.

The radial total correlation energy density, calculated with both the positive-definite, $E^+_{c,\,\text{rad}}(r)$, and Schr{\"o}dinger, $E^s_{c,\,\text{rad}}(r)$, kinetic energy densities for the \ce{He} atom are compared to the total potential correlation energy density in Figure \ref{fig:hetv}.  Both forms of total correlation energy densities differ from the simpler potential correlation energy density, albeit in different manners.  The Schr{\"o}dinger total correlation energy density has a sharp peak at small $r$, while the potential correlation energy density is negative for all $r$.  The positive-definite total correlation energy density has two dips while the potential correlation energy density has a single dip.

In the case of \ce{Ne} and \ce{Ar}, Figures \ref{fig:netv} and \ref{fig:artv}, a similar trend is observed.  It is seen that, for these many-electron atoms, the agreement between the potential correlation energy density and the total correlation energy densities improves over that seen for the \ce{He} atom.  However, differences remain. In general they are summarized as: there is relative agreement between the magnitude of the features of the potential correlation energy and the Schr{\"o}dinger total correlation energy, whereas the actual features ({\it i.e.} dips and peaks) of the potential correlation energy agree better with the smoother positive-definite total correlation energy density.  Of course, in order to satisfy the Virial Theorem locally, the energy densities should not agree in magnitude but differ by a factor of 2.  Therefore, for these particular definitions of the total correlation energy it is the cancellation of the oscillations (positive and negative) that allow the energy densities to satisfy the overall (integrated) Virial Theorem.

\section{Conclusions}
\label{sec:conc}

Atomic radial correlation energy densities were computed for the first 17 multiple-electron atoms of the periodic table.  The correlation energy densities were decomposed into components corresponding to the terms of the electronic Hamiltonian; nucleus-electron potential, kinetic and electron-electron potential energy densities.

Considering the potential application of the calculation and analysis of these atomic radial correlation energy densities to molecules and molecular interactions, their accuracy with respect to integration grid was assessed.  The assessment included an analysis of both the HF and exFCI energy density components along with the correlation energy densities (the difference).  It was found that for the energy components, other than the kinetic, the SG-1 grid is sufficient for accuracy of at least 1 kJ mol$^{-1}$.  However, for kinetic energy and a more subtle analysis of energetics, the SG-2 grid would be required.  In the case  of the correlation energy components, partly due to a cancellation of error, the results using the SG-2 grid are highly accurate.

The effect of electron correlation on the nucleus-electron potential energy is explained by the change in the one-electron density.  This typically involves an accumulation of density near the nucleus and at large nucleus-electron distance, or the opposite, a depletion near and far from the nucleus with a build up at mid-range.  Which scenario occurs can depend on whether the electrons are $\alpha$ or $\beta$-spin (see the \ce{Li} atom).  As electrons are added to new subshells, new regions of depletion and addition of electron density [or positive and negative $V^{\text{ne}}_{c,\text{rad}}(r)$] appear and oscillate with increasing $r$. The larger the atomic number the more $V^{\text{ne},\alpha}_{c,\text{rad}}(r)$ and $V^{\text{ne},\beta}_{c,\text{rad}}(r)$ resemble each other.

Both the, Laplacian based, Schr{\"o}dinger kinetic correlation energy density and the, gradient based, positive-definite kinetic correlation energy density were analyzed and compared.  As expected, the Schr{\"o}dinger kinetic correlation energy density has more features (peaks and dips) compared to the ``smoother" positive-definite kinetic correlation energy density.  In general, the Schr{\"o}dinger kinetic correlation energy density is mainly positive and has a smaller amplitude than the more oscillatory positive-definite kinetic correlation energy density.

The spin-pair-resolved electron-electron potential correlation energy densities of atoms from \ce{He} to \ce{Ar} were analyzed.  It is clear from the amplitude of the correlation energy densities that, from left to right on the periodic table, the amount of correlation per electron pair decreases.  There is also a clear separation of the effects of correlation on the electrons of each principal shell ($n=1,2,3$).  It was also found that for the \ce{Be} atom and beyond, there are regions of increased repulsion (positive correlation energy density) due to correlation.  The analysis also highlighted the difference of the observed effects of correlation on opposite-spin electrons depending the spin of the reference electron ($\alpha$ or $\beta$).

Finally, the locality of the Virial Theorem was investigated with both of the present definitions of kinetic energy density.  While the potential energy and total energy densities shared some similar features ({\it e.g.}~concavity and convexity) at similar locations, there is quite a discrepancy between the two.

The approach employed here to analyze the radial correlation energy density components of isolated atoms can easily be extended to molecules and molecular clusters through the use of atomic weights.  The energy densities of individual atoms, as seen here, can be quite complex and therefore the present analysis will serve as useful reference for future studies of correlation in molecular systems.  Of course, as the number of atoms grows, the particular form of CI, or some alternative post-HF, wave function that is calculated will need to be tailored in consideration of computational cost.  However, as long as a 1-RDM and 2-RDM can be extracted then the above approach may be employed.

\begin{acknowledgments}
R.A.P. is grateful to the Natural Sciences and Engineering Council of Canada (NSERC) for financial support. A.A.A thanks the School of Graduate Studies of Memorial University of Newfoundland and Chen Graduate Scholarship for funding. JWH thanks NSERC for a Discovery Grant, the Digital Research Alliance of Canada for computing resources and the Discovery Institute for Computation and Synthesis for useful consultations.
\end{acknowledgments}

\section*{Data Availability}
The data that support the findings of this study are available in the following GitLab repository \url{https://gitlab.com/qctsl/atomicecden.git}.

\bibliography{atomicEcden}

\clearpage

\begin{widetext}
\begin{center}
{\textbf{\large Supplementary Material: Atomic Radial Correlation Energy Density Components}}
\end{center}
\end{widetext}

\section*{S.I CIPSI wave functions}
The extrapolated full CI (exFCI) and unextrapolated CIPSI, energies, along with the number of determinants used in the CIPSI expansion, for the atoms \ce{He} to \ce{Ar}, are reported in Table \ref{tab:wf}. Each exFCI calculation was run using the natural orbitals obtained from a previous, smaller, exFCI calculation with a maximum of $10^4$ determinants. The final exFCI calculations were terminated when the perturbative correction fell below $10^{-4}$ hartrees, or the CIPSI expansion increased beyond $10^6$ determinants.
\FloatBarrier
\begin{table}
	\renewcommand\thetable{S1}
	\caption{\label{tab:wf} Extrapolated and unextrapolated CIPSI energies, and the number of determinants for the wave functions of helium to argon.}
	\begin{ruledtabular}
	\begin{tabular}{lrrr}
	Atom	& $E_\text{exFCI}$	& $E_\text{CIPSI}$	& $N_\text{det}$	\\
	\hline
	\ce{He}		& -2.900836	& -2.900836	&  55		\\
	\ce{Li}		& -7.474553	& -7.474499	&  318		\\
	\ce{Be}		& -14.662544	& -14.662512	&  3428		\\
	\ce{B}		& -24.643754	& -24.643687	& 28345		\\
	\ce{C}		& -37.826986	& -37.826911	& 141676	\\
	\ce{N}		& -54.561284	& -54.561214	& 1758265	\\
	\ce{O}		& -75.018177	& -75.018108	& 1420974	\\
	\ce{F}		& -99.664278	& -99.664172	& 1066172	\\
	\ce{Ne}		& -128.848690	& -128.848602	& 1775214	\\
	\ce{Na}		& -162.123871	& -162.123813	& 1301836	\\
	\ce{Mg}		& -199.927039	& -199.926600	& 1157774	\\
	\ce{Al}		& -242.219980	& -242.219133	& 1947753	\\
	\ce{Si}		& -289.222654	& -289.220435	& 1179100	\\
	\ce{P}		& -341.108144	& -341.105566	& 1693107	\\
	\ce{S}		& -397.933771	& -397.929715	& 1496069	\\
	\ce{Cl}		& -459.949679	& -459.944105	& 1067357	\\
	\ce{Ar}		& -527.321338	& -527.316386	& 1329503	\\
	\end{tabular}
	\end{ruledtabular}
\end{table}
\FloatBarrier

\section*{S.II Correlation energy components}
The analytically calculated, spin-resolved, one and two-electron correlation energy components for the atoms helium to argon are reported in Tables \ref{tab:ec1} and \ref{tab:ec2}.
\begin{table}
	\renewcommand\thetable{S2}
	\caption{\label{tab:ec1} One-electron correlation energy components of the atoms helium to argon.}
	\begin{ruledtabular}
	\begin{tabular}{lrrrr}
	Atom	& $T^\alpha_c$	& $T^\beta_c$	& $V^{\text{ne},\alpha}_c$	& $V^{\text{ne},\beta}_c$	\\
	\hline
	\ce{He}	& 0.018336	& 0.018336	& 0.000481	& 0.000481 \\
	\ce{Li}	& 0.036510	& 0.003154	&-0.022237	& 0.019154 \\
	\ce{Be}	& 0.043152	& 0.043152	&-0.032920	&-0.032920 \\
	\ce{B}	& 0.066708	& 0.043123	&-0.049316	&-0.008001 \\
	\ce{C}	& 0.109301	& 0.021855	&-0.085761	& 0.049699 \\
	\ce{N}	& 0.178611	&-0.025888	&-0.154097	& 0.146025 \\
	\ce{O}	& 0.359399	&-0.157149	&-0.439662	& 0.479030 \\
	\ce{F}	& 0.336603	&-0.088089	&-0.341311	& 0.419699 \\
	\ce{Ne}	& 0.148130	& 0.148141	& 0.060433	& 0.060411 \\
	\ce{Na}	& 0.095687	& 0.081636	& 0.041193	& 0.084640 \\
	\ce{Mg}	& 0.115915	& 0.115913	&-0.014781	&-0.014780 \\
	\ce{Al}	& 0.146677	& 0.092152	&-0.076187	& 0.046955 \\
	\ce{Si}	& 0.204521	& 0.049027	&-0.167811	& 0.118828 \\
	\ce{P}	& 0.288335	&-0.014498	&-0.273722	& 0.212570 \\
	\ce{S}	& 0.362705	&-0.046147	&-0.459930	& 0.390897 \\
	\ce{Cl}	& 0.310533	& 0.037647	&-0.324066	& 0.245775 \\
	\ce{Ar}	& 0.189794	& 0.189793	&-0.037538	&-0.037533 \\
	\end{tabular}
	\end{ruledtabular}
\end{table}
\begin{table}
	\renewcommand\thetable{S3}
	\caption{\label{tab:ec2} Two-electron correlation energy components of the atoms helium to argon.}
	\begin{ruledtabular}
	\begin{tabular}{lrrr}
	Atom	& $V^{\text{ee},\alpha\beta}_c$	& $V^{\text{ee},\alpha\alpha}_c$ & $V^{\text{ee},\beta\beta}_c$	\\
	\hline
	\ce{He}	&-0.038624	& 0		& 0 \\
	\ce{Li}	&-0.039702	& 0.000977	& 0 \\
	\ce{Be}	&-0.064521	& 0.009475	& 0.009475 \\
	\ce{B}	&-0.088538	& 0.006109	& 0.002904 \\
	\ce{C}	&-0.111447	&-0.005493	&-0.006794 \\
	\ce{N}	&-0.132730	&-0.023625	&-0.019000 \\
	\ce{O}	&-0.201238	& 0.036377	&-0.089641 \\
	\ce{F}	&-0.235363	&-0.001792	&-0.117882 \\
	\ce{Ne}	&-0.272694	&-0.093113	&-0.093108 \\
	\ce{Na}	&-0.211173	&-0.067371	&-0.078533 \\
	\ce{Mg}	&-0.208498	&-0.048448	&-0.048448 \\
	\ce{Al}	&-0.224393	&-0.038191	&-0.064922 \\
	\ce{Si}	&-0.231351	&-0.029464	&-0.076902 \\
	\ce{P}	&-0.236884	&-0.030493	&-0.090856 \\
	\ce{S}	&-0.271322	& 0.028087	&-0.155148 \\
	\ce{Cl}	&-0.288105	&-0.013366	&-0.138958 \\
	\ce{Ar}	&-0.307212	&-0.091942	&-0.091944 
	\end{tabular}
	\end{ruledtabular}
\end{table}

\end{document}